\documentclass[useAMS,letterpaper,usenatbib,usegraphicx]{mn2e}

\newcommand{\snd}{secondary}

\newcommand{\zdg}{$^{\circ}$}

\newcommand{\ks}{km s$^{-1}$}

\newcommand{\mrs}{maximum redshift}
\newcommand{\wtd}{white dwarf}

\usepackage{times}
\usepackage{mathptm}
\begin{document}


\title[Discovery of a Persistent QPO in IP TX Col]{The Discovery of a Persistent Quasi-Periodic Oscillation in the Intermediate Polar TX Col}
\author[N. Mhlahlo et al.]{N. Mhlahlo$^{1}$\thanks{E-mail:
nceba@maia.saao.ac.za}, D.A.H. Buckley$^{2}$, V.S. Dhillon$^{3}$, S.B. Potter$^{2}$, B. Warner$^{1}$, P. Woudt$^{1}$, \newauthor G. Bolt$^{4}$, J. McCormick$^{5}$, R. Rea$^{6}$, Denis J. Sullivan$^{7}$ and F. Velhuis$^{5}$  \\
$^{1}$Astronomy Department, University of Cape Town, Rondebosch 7700, Cape Town, South Africa \\
$^{2}$South African Astronomical Observatory, Observatory 7935, Cape Town, South Africa \\
$^{3}$Physics and Astronomy Department, University of Sheffield, Sheffield, S3 7RH, UK  \\
$^{4}$Centre for Backyard Astrophysics (Perth), 295 Camberwarra Drive, Craigie, Western Australia 6025, Australia  \\
$^{5}$Centre for Backyard Astrophysics (Pakuranga), Farm Cove Observatory, 2/24 Rapallo Place, Farm Cove, Pakuranga, Auckland, New Zealand \\
$^{6}$Centre for Backyard Astrophysics (Nelson), Regent Lane Observatory, Nelson, New Zealand \\
$^{7}$Department of Physics, Victoria University of Wellington, Wellington, New Zealand.}


\maketitle

\label{firstpage}

\begin{abstract}
We report on the detection of an $\sim$5900 s quasi-periodic variation in the extensive photometry of TX Col spanning 12 years.  We discuss five different models to explain this period.  We favour a mechanism where the quasi-periodic variation results from the beating of the Keplerian frequency of the `blobs' orbiting in the outer accretion disc with the spin frequency, and from modulated accretion of these `blobs' taking place in a shocked region near the disc/magnetosphere boundary.    \\
\end{abstract}


\section{\bf Introduction}
\label{sec:txintro}

TX Col was first discovered as an X-ray source (1H0542-407) in the HEAO-1 all-sky survey.  X-ray (EXOSAT) and optical observations~\citep{tuo86,buc89} established this system as a new Intermediate Polar (IP), a subclass of magnetic cataclysmic variable stars (mCVs) where the \wtd\ is in asynchronous rotation with the orbital motion of the system.  A white dwarf rotation period of $\sim$1911 s and an orbital period of $\sim$ 5.7 hr were determined from a combination of radial velocity, X-ray and optical intensity modulations. 
TX Col showed very hard X-ray spectra (kT $\ge$ 10 keV), with the hard X-rays modulated strongly at the beat period (2106 s).  The hard X-rays are thought to result due to a strong shock forming above the white dwarf surface where the accreted material is heated to high temperatures ($\sim$10$^{8}$ K) \citep{nor97} and are reflected and reprocessed in regions fixed in the binary frame (the bright spot or the \snd), producing the beat period.  

Observed changes in the amplitude and power spectra of the optical light curves of TX Col over a long period of time (1989-2002), signifying variations in its accretion behaviour, have sparked a debate concerning the exact accretion mode in TX Col: whether or not accretion occurs via a disc, directly from the accretion stream or some combination of both (known as disc-overflow accretion; \cite{nor97}).  
  
The detection by \cite{tuo86} and \cite{buc89} of the beat period in the photometry and X-rays was indicative of strong disc-overflow, stream-fed accretion or even reprocessing from regions that are fixed in the rotating frame of the binary. 

Later optical photometry in 1989 (Buckley \& Sullivan 1992) showed a persistent periodicity at $\sim$1054 s, exactly half the previously observed beat period of $\sim$2106 s.  This 1054-s harmonic was not seen in the previously published photometry and was attributed to reprocessing of X-rays from both magnetic poles in regions fixed in the orbital phase.  This could also be due to direct or overflowing stream of material flipping between the two magnetic poles of the \wtd.

Further optical photometry of TX Col was obtained at the South African Astronomical Observatory (SAAO), Cerro Tololo Inter-American Observatory (CTIO) and the Mt. John University Observatory (MJUO) in 1994 \citep{buc94}, which no longer showed either the beat period or its harmonic, but instead revealed a strong period near 6000 s and other quasi-periodic light variations at similar low frequencies.  
\begin{table}
\begin{center}
\small
\caption{\small Observation table of the 2002 photometry of TX Col obtained over a period of a month at the SAAO and by the CBA stations in New Zealand and Australia.}  \label{tab:observ1}
\begin{tabular}{|c|c|c|c|c|}  \hline
\multicolumn{1}{|c|}{\textbf{Observing}} &
\multicolumn{1}{|c|}{\textbf{Place}} &
\multicolumn{1}{|c|}{\textbf{HJD (Start)}} &
\multicolumn{1}{c|}{\textbf{Exp.}} &   
\multicolumn{1}{c|}{\textbf{Length}}  \\  
  \textbf{Date} &  & 2450000+  &   \textbf{Time (s)} & (h)   \\ \hline
 02 Jan 02 &CBA: Pakuranga& 2276.8955 &35 & 6.7  \\  
 05 Jan 02 &CBA: Perth& 2280.0609   &54 & 4.1 \\   
 06 Jan 02 &CBA: Pakuranga& 2280.8871 &35 & 3.9 \\  
 10 Jan 02 &CBA: Perth& 2285.0623   &54 & 5.6 \\   
 10 Jan 02 &CBA: Perth& 2286.0304   &54 & 3.2 \\   
 15 Jan 02 & SAAO & 2290.4031 &20 & 5.0 \\  
 16 Jan 02 &CBA: Pakuranga& 2290.8856 &35 & 6.3 \\  
 16 Jan 02 & SAAO & 2291.3113 &20 & 3.3 \\   
 18 Jan 02 & SAAO & 2293.2848 &20 & 8.0 \\  
 19 Jan 02 &CBA: Perth& 2294.0340   &54 & 6.4 \\
 19 Jan 02 & SAAO & 2294.2872 &20 & 8.0 \\     
 20 Jan 02 &CBA: Pakuranga& 2294.8728 &35 & 7.1 \\   
 20 Jan 02 &CBA: Perth& 2295.0630   &54 & 5.7 \\  
 20 Jan 02 & SAAO & 2295.2811 &20 & 8.2  \\  
 21 Jan 02 & SAAO & 2296.3296 &20 & 6.7  \\  
 22 Jan 02 &CBA: Pakuranga& 2296.8773 &35 & 7.1 \\ 
 22 Jan 02 & SAAO & 2297.3502 &20 & 6.0  \\   
 23 Jan 02 &CBA: Nelson& 2297.8740 &35 & 5.9 \\
 23 Jan 02 &CBA: Pakuranga& 297.9040  &35 & 1.7 \\
 23 Jan 02 & SAAO & 2298.2809 &20 & 5.0  \\  
 25 Jan 02 & SAAO & 2299.2905 &20 & 4.3  \\    
 26 Jan 02 & SAAO & 2301.4193 &20 & 4.4  \\        
 27 Jan 02 &CBA: Perth& 2302.0760   &54 & 4.9 \\
 27 Jan 02 & SAAO & 2302.2681 &20 & 8.1  \\   
 28 Jan 02 &New Zealand& 2302.8933 &35 & 3.7 \\  
 29 Jan 02 &CBA: Pakuranga& 2303.8625 &35 & 6.9 \\  
 29 Jan 02 &CBA: Nelson& 2303.8877 &35 & 5.5 \\   
 29 Jan 02 & SAAO & 2304.2765 &20 & 7.6  \\   
 30 Jan 02 &CBA: Nelson& 2304.8810 &35 & 7.4 \\  
 31 Jan 02 &CBA: Perth& 2306.0758   &54 & 4.6 \\  
 31 Jan 02 & SAAO & 2306.2631 &20 & 4.8 \\  
 01 Feb 02 &CBA: Perth& 2307.0115   &54 & 5.7 \\
 01 Feb 02 & SAAO & 2307.2613 &20 & 7.8 \\  
 02 Feb 02 & SAAO & 2308.2611 &20 & 7.8 \\      
 03 Feb 02 &CBA: Nelson& 2308.9022 &40 & 5.3 \\
  03 Feb 02 & SAAO & 2309.2672 &20 & 6.7 \\  
 04 Feb 02 & SAAO & 2310.2608 &20 & 7.8 \\   
 06 Feb 02 &CBA: Nelson& 2311.8649 &40 & 6.3 \\ \hline 
\end{tabular}   		       
\end{center}			      
\end{table} 			      
\begin{table}
\begin{center}
\small
\caption{\small Observation table for photometry obtained from near the end of 1989 to near the end of 1990, in 1991 and 1994 at SAAO and at MJUO.}  \label{tab:observ2}
\begin{tabular}{|c|c|c|c|c|}  \hline
\multicolumn{1}{|c|}{\textbf{Observ.}} &
\multicolumn{1}{|c|}{\textbf{Place}} &
\multicolumn{1}{|c|}{\textbf{HJD (Start)}} &
\multicolumn{1}{c|}{\textbf{Exp.}} &  
\multicolumn{1}{c|}{\textbf{Length}}  \\  
  \textbf{Date} &  & 2440000+  &  \textbf{time (s)} & (h)   \\ \hline
 26 Nov 89 &SAAO& 7857.2898    & 10 & 7.2 \\  
 28 Nov 89 &MJUO & 7858.9643   & 10 & 4.3 \\   
 29 Nov 89 &MJUO  & 7859.9037  & 10 & 4.9 \\  
 18 Jan 90 &SAAO & 7910.3101   & 10 & 1.2 \\   
 19 Jan 90 &SAAO  & 7911.3036  & 10 & 1.8 \\  
 16 Sep 90 &SAAO  & 8150.5609  & 5  & 2.2 \\  
 21 Sep 90 &SAAO  & 8156.4647  & 10 & 4.3 \\   
 09 Nov 90 &SAAO  & 8205.3513  & 10 & 6.0 \\  
 12 Nov 90 &SAAO  & 8208.3624  & 10 & 5.7 \\   
 20 Nov 90 &SAAO  & 8216.4194  & 20 & 4.2 \\  
 18 Dec 90 &SAAO  & 8244.3007  & 10 & 7.0 \\  
 19 Dec 90 &SAAO  & 8245.3192  & 10 & 6.6 \\   
 20 Dec 90 &SAAO  & 8246.2972  & 10 & 7.0 \\  
 21 Dec 90 &SAAO  & 8247.2986  & 10 & 6.7 \\  
 22 Dec 90 &SAAO  & 8248.2979  & 10 & 7.1 \\   
 23 Dec 90 &SAAO  & 8249.2983  & 10 & 7.2 \\  
 24 Dec 90 &SAAO  & 8250.3812  & 10 & 1.0 \\  
 24 Dec 90 &SAAO  & 8250.4514  & 10 & 3.5 \\   
 10 Apr 91 &SAAO& 8357.2506 &  10 & 3.4 \\ 
 12 Apr 91 &SAAO& 8359.2727 &  10 & 3.0 \\  
 13 Apr 91 &SAAO& 8360.2279 &  10 & 2.5 \\ 
 18 Apr 91 &SAAO& 8365.2443 &  10 & 2.4 \\  
 31 Oct 91 &SAAO& 8561.3939 &  10 & 3.3 \\   
 01 Nov 91 &SAAO& 8562.3618 &  10 & 5.9 \\  
 02 Nov 91 &SAAO& 8563.3508 &  10 & 6.4 \\  
 03 Nov 91 &SAAO& 8564.3396 &  10 & 6.2 \\   
 04 Nov 91 &SAAO& 8565.3481 &  10 & 6.4 \\  
 05 Nov 91 &SAAO& 8566.3897 &  10 & 3.2 \\   
 08 Nov 91 &SAAO& 8569.4055 &  10 & 4.5 \\  
 09 Nov 91 &SAAO& 8570.3421 &  10 & 4.0 \\  
 11 Nov 91 &SAAO& 8572.3390 &  10 & 4.0 \\   
 09 Dec 91 &SAAO& 8599.9170 &  10 & 5.6 \\  
 10 Dec 91 &SAAO& 8600.9501 &  10 & 4.7 \\ 
 10 Jan 94 &MJUO& 9363.0505  &  5  & 2.2 \\  
 11 Jan 94 &MJUO & 9363.9155 &  5  & 5.7 \\  
 11 Jan 94 &SAAO& 9364.3142 &  5    & 5.2 \\    
 12 Jan 94 &MJUO & 9364.9178 &  5  & 4.2 \\   
 13 Jan 94 &SAAO& 9366.3508 &  5    & 5.2 \\  
 14 Jan 94 &SAAO& 9367.3141  & 5    & 3.8 \\  
 14 Jan 94 &SAAO& 9367.4797  & 5    & 1.3 \\   
 15 Jan 94 &SAAO& 9368.3107  & 5    & 6.1 \\  
 15 Jan 94 &CTIO& 9367.5792   & 10  & 2.1 \\  
 16 Jan 94 &MJUO & 9368.9178  & 5   & 3.9 \\    
 16 Jan 94 &CTIO & 9368.5783  & 10  & 2.4 \\   
 16 Jan 94 &SAAO& 9369.2957  & 5    & 5.4 \\  
 17 Jan 94 &CTIO & 9369.5518  & 10  & 2.2 \\  
 17 Jan 94 &SAAO& 9370.3225  & 5    & 5.0 \\   
 18 Jan 94 &CTIO & 9370.5740  & 10  & 1.4 \\ \hline  
\end{tabular}   
\end{center}
\end{table} 
Our 2002 observations reported here, together with those obtained by the CBA (Center for Backyard Astrophysics), show that TX Col power spectra were dominated by high-amplitude quasi-periodic light variations in 2002.  A prominent quasi-periodic oscillation (QPO) period at $\sim$5900 s ($\sim$170 $\mu$ Hz) was detected, the same period as detected in the data of 1990 and 1994. 

The purpose of this study is to investigate the origin/cause of this oscillation.  We start by presenting the photometry of TX Col in Section~\ref{tx:phot}, and in Section~\ref{sec:dfts} we analyse the entire data set, i.e. our 2002 data and the archival data from 1989 to 1994.  The analysis of the QPO period is done in Section~\ref{sec:qpo} and in Section~\ref{dis:alldisc} we discuss and interpret the results.

\section{\bf Photometric Observations}
\label{tx:phot}
The optical photometry of TX Col was obtained at SAAO in January 2002 using the 1.0-m telescope at Sutherland with UCT CCD photometer in frame-transfer mode using B and I filters.  Additional photometry was obtained by the CBA group nearly at the same period.  The archival data obtained from SAAO, MJUO and CTIO, from 1989 - 1994, were retrieved and analysed alongside the CBA and our 2002 photometry.  No filters were used for the CBA and the archival data.

The photometry was grouped and analysed in three sections:  the SAAO and CBA data sets combined (hereafter the 2002 combined photometry), the 1989, 1990, 1991 and 1994 data sets combined (hereafter the archival photometry) and the 2002 combined data together with the archival photometry combined (hereafter the 1989-2002 combined photometry). 
\subsection{The 2002 Combined Data Reduction}
\label{sec:photanli}
For the SAAO observations the integration times were 20 s. 
Sky flatfields were taken at twilight throughout the observation week.
The observation period was nearly three weeks and 6349 B-band images in total were taken.  
The data were reduced using the Dophot program \citep{mat89}.  
\begin{figure*}
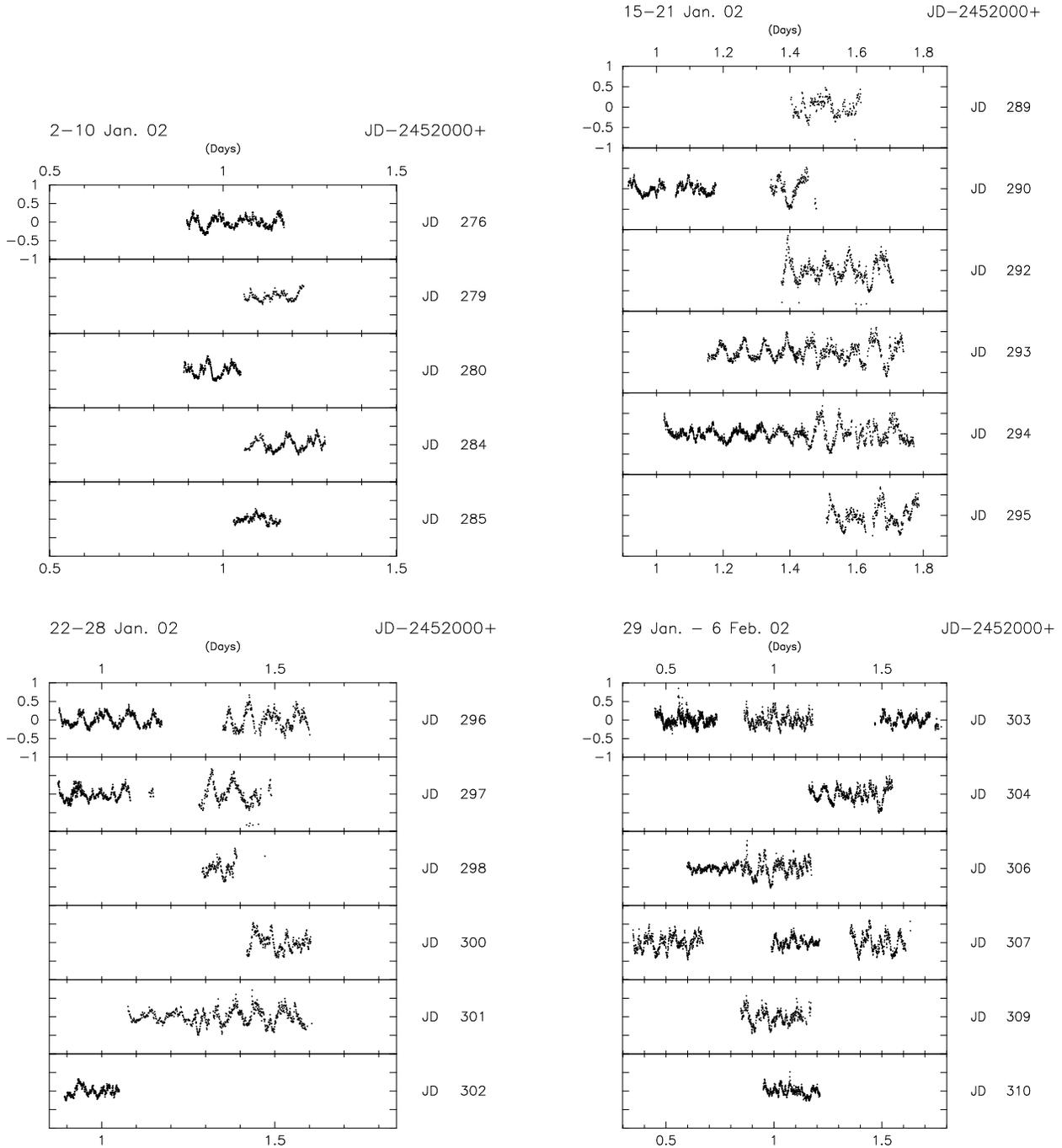

\begin{center}
\small
\includegraphics[height=90mm,width=91mm]{tx_jan_allx4flc1ai.eps}
\includegraphics[height=90mm,width=85mm]{tx_jan_allx4flc2a.eps}
\includegraphics[height=90mm,width=91mm]{tx_jan_allx4flc3a.eps}
\includegraphics[height=90mm,width=85mm]{tx_jan_allx4flc4a.eps}
\caption{\small TX Col light curves obtained from 2 January to 6 February 2002 at SAAO and by CBA groups in New Zealand and Australia.  The ordinates are intensity measurements with the mean subtracted and normalized.  1 on the x-axis corresponds to the Julian Day (JD) value shown to the right of the plots.  Time is plotted on the abscissa, i.e. the values on the horizontal axis add or subtract to the JDs, depending on whether the data points lie before or after the 1 on the x-axis.}
\label{l:lcplotpa}
\end{center}
\end{figure*}
The CBA photometry was acquired during the period 02 January - 06 February 2002 spanning the entire SAAO campaign.  CBA observers in Australia (Perth) and New Zealand (Pakuranga and Nelson) participated in the 2002 campaign.  A 0.35-m Schmidt-Cassegrain telescope with an SBIG (Santa Barbara Instruments Group) ST6 CCD camera (CBA: Nelson),
Meade LX200 10" f/10 with an SBIG ST7e CCD camera (CBA: Pakuranga) and 10" f6.3 LX200 SBIG ST7 CCD camera (CBA: Perth) were used.  
\begin{figure*}
\begin{center}
\includegraphics[angle=0,width=140mm]{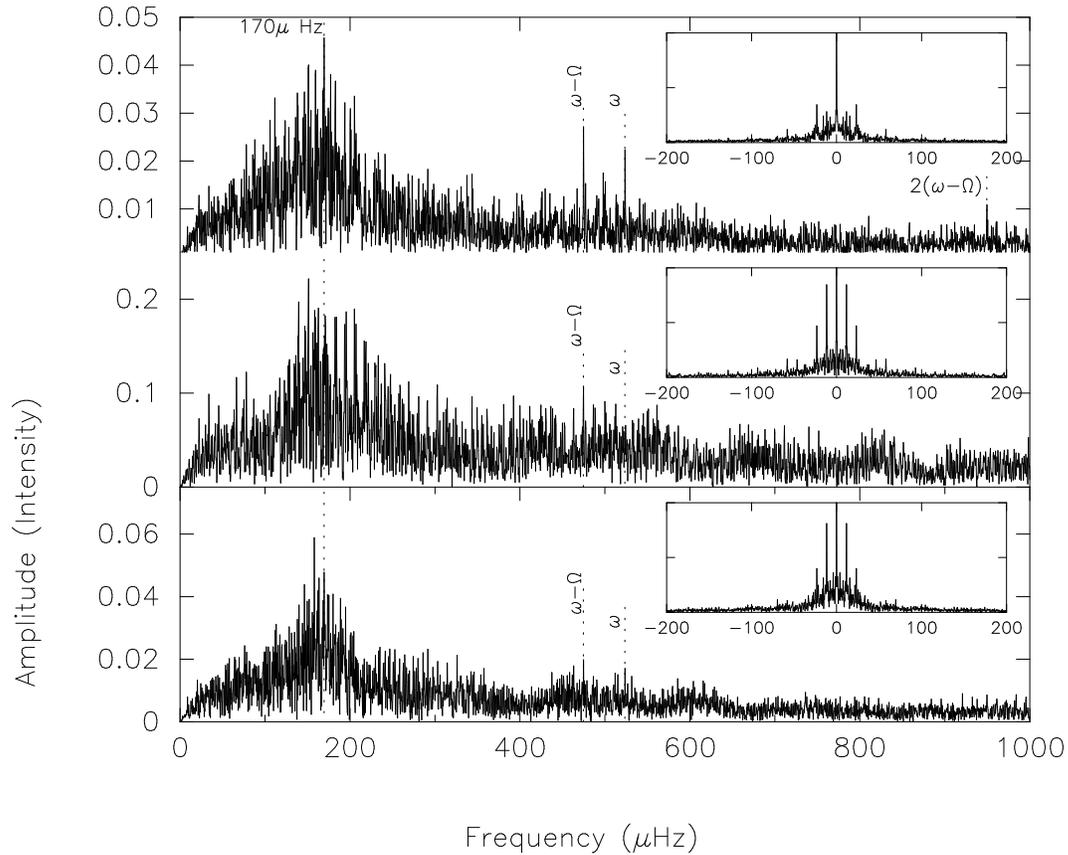}
\caption{\small Amplitude spectra of the CBA data (bottom panel), the SAAO data (middle panel) and the 2002 combined data (top panel).  The inserts shown are the window spectra for each data set.}
\label{l:photplot}
\end{center}
\end{figure*}
Table~\ref{tab:observ1} shows the observation logs.
  
The CBA and the SAAO photometry were combined before the analyses.  Before analysis of the two data sets, all the data were converted from magnitude scale into relative intensity scale and the mean for each set subtracted and used to normalise.  This was done because the data were obtained from different instruments in different scales.  A sample of the normalised light curves of the 2002 combined photometry are shown in Figure~\ref{l:lcplotpa}.  The SAAO data show excursions (large variations in amplitude) which are not seen in the CBA data.  They are possibly due to the effect of the filter.  

\subsection{The Archival Data and The 1989-2002 Combined Photometry}  
\label{sec:allcomb}
The white-light archival photometry obtained in November 1989; January, September, November and December 1990; April, November and December 1991 and in January 1994 at the SAAO, MJUO and CTIO was also analysed.  The SAAO 0.75 and 1.0-m telescopes were used with the UCT photometer employing a photomultiplier.  For the MJUO observations a two channel photomultiplier photometer attached to the McClellan1.0-m telescopes was used.  Table~\ref{tab:observ2} shows the observations.

\section{Period Analysis}
\label{sec:dfts} 
Discrete Fourier Transforms (DFTs) were produced \citep{kur85} to reveal the periodicities in the data.  The results are displayed in Figure~\ref{l:photplot} for the 2002 combined photometry. The spin and the beat frequencies are detected at $\omega=523.636 \pm0.019$ $\mu$Hz and $\omega-\Omega=474.788 \pm0.014$ $\mu$Hz, respectively.  
\begin{table}
\begin{center}
\caption{\small Measured amplitudes and phases at QPO peak maximum obtained from least squares fitting of the 169.56 $\mu$Hz QPO to the 2002 combined photometry from 15-23 Jan 2002.  The first data point, HJD=2452291.311340, of 16 Jan (SAAO) was used as a phase reference point.  (NZ) and (Aust.) denote data obtained in New Zealand and Australia, respectively, by the CBA group.  Norm. Intensity refers to normalised intensity.} 
\label{tab:02splqpopar}
\small
\begin{tabular}{|c|c|c|c|}  \hline
\multicolumn{1}{|c|}{\textbf{Date}} &
\multicolumn{1}{|c|}{\textbf{Place}} &
\multicolumn{1}{c|}{\textbf{Amplitude}} &
\multicolumn{1}{c|}{\textbf{Phase of Max.}}                   \\
(2002)  &   &  (Norm. Intensity)      & (cycles) \\ \hline
 15 Jan. & SAAO& $0.04\pm0.02$ &   $0.1\pm0.09$   \\ 
 16 Jan. & CBA (NZ) & $0.03\pm0.01$ &    $0.97\pm0.04$   \\  
 16 Jan. & SAAO & $0.30\pm0.01$ &  $0.39\pm0.01$   \\  
 18 Jan. & SAAO & $0.13\pm0.02$ &  $0.13\pm0.02$   \\   
 19 Jan. & CBA (Aust.) & $0.20\pm0.01$ & $0.28\pm0.01$   \\   
 19 Jan. & SAAO & $0.22\pm0.01$ &  $0.317\pm0.01$  \\  
 20 Jan. & CBA (NZ) & $0.10\pm0.01$ &    $0.35\pm0.01$ \\  
 20 Jan. & CBA (Aust.) & $0.10\pm0.01$ & $0.37\pm0.01$   \\ 
 20 Jan. & SAAO & $0.14\pm0.01$ &  $0.1\pm0.02$    \\ 
 21 Jan. & SAAO & $0.07\pm0.02$ &  $0.21\pm0.04$   \\  
 22 Jan. & CBA (NZ) & $0.15\pm0.01$ &   $0.48\pm0.01$  \\  
 22 Jan. & SAAO & $0.21\pm0.01$ &  $0.60\pm0.01$   \\  
 23 Jan. & CBA (NZ) & $0.11\pm0.01$ &    $1.00\pm0.01$   \\   
 23 Jan. & SAAO & $0.22\pm0.02$ &  $0.62\pm0.01$    \\ \hline  
\end{tabular}   
\end{center}
\end{table} 
\begin{figure*}
\begin{center}
\includegraphics[width=177mm]{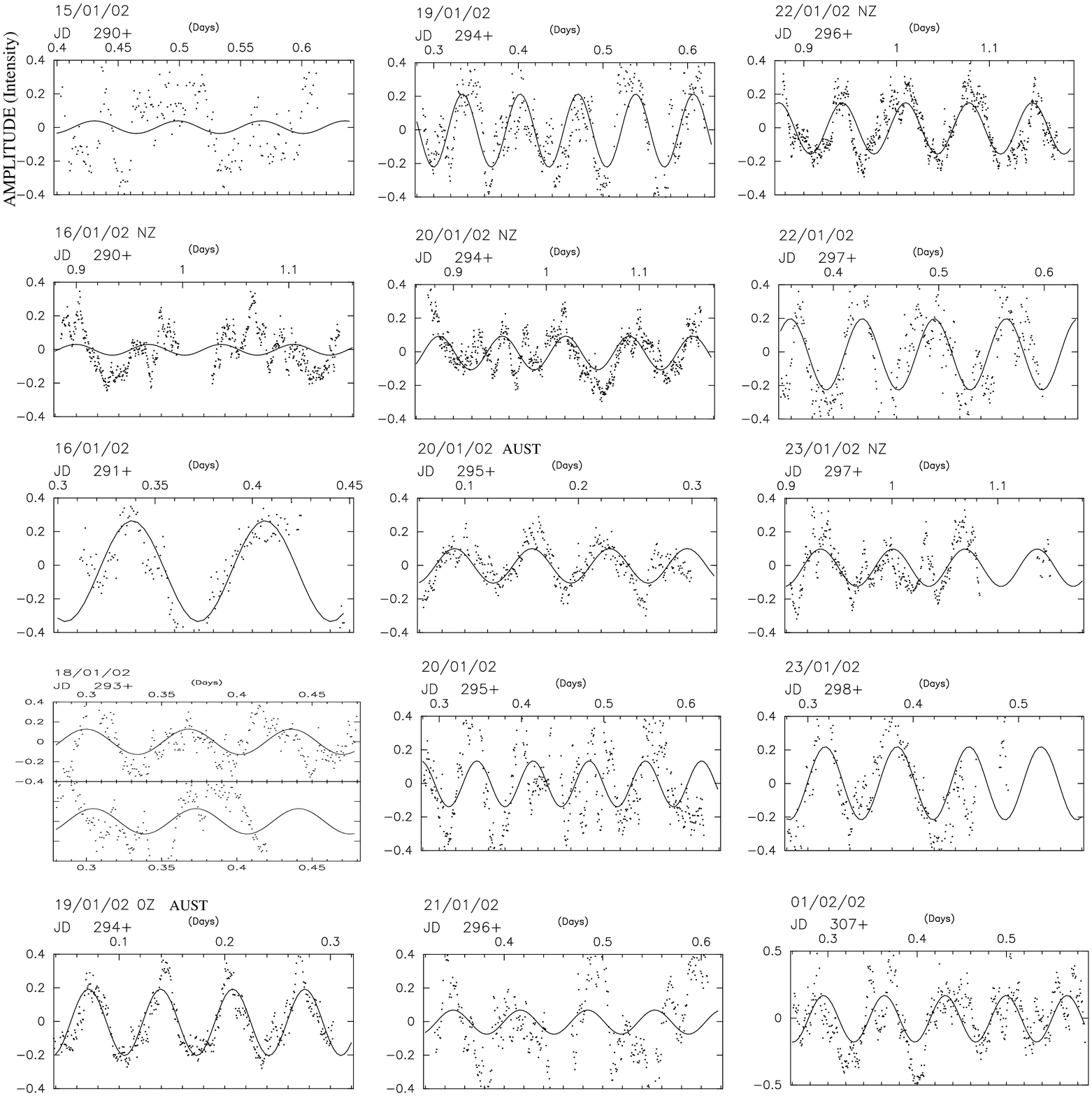}
\caption{\small Sample light curves of the 2002 combined photometry on consecutive nights.  NZ and AUST stand for CBA stations in New Zealand and in Australia, respectively, and the rest of the panels are runs obtained at SAAO.  The QPO periodicity was fitted to the data as represented by a solid line.  A strong variation at the QPO period can be seen during a number of runs, more especially on the 19 January and the 22 January.  The JDs run from 2452290 to 2452307.}
\label{l:archivespl}
\end{center}
\end{figure*}
\begin{figure*}
\begin{center}
\includegraphics[width=120mm]{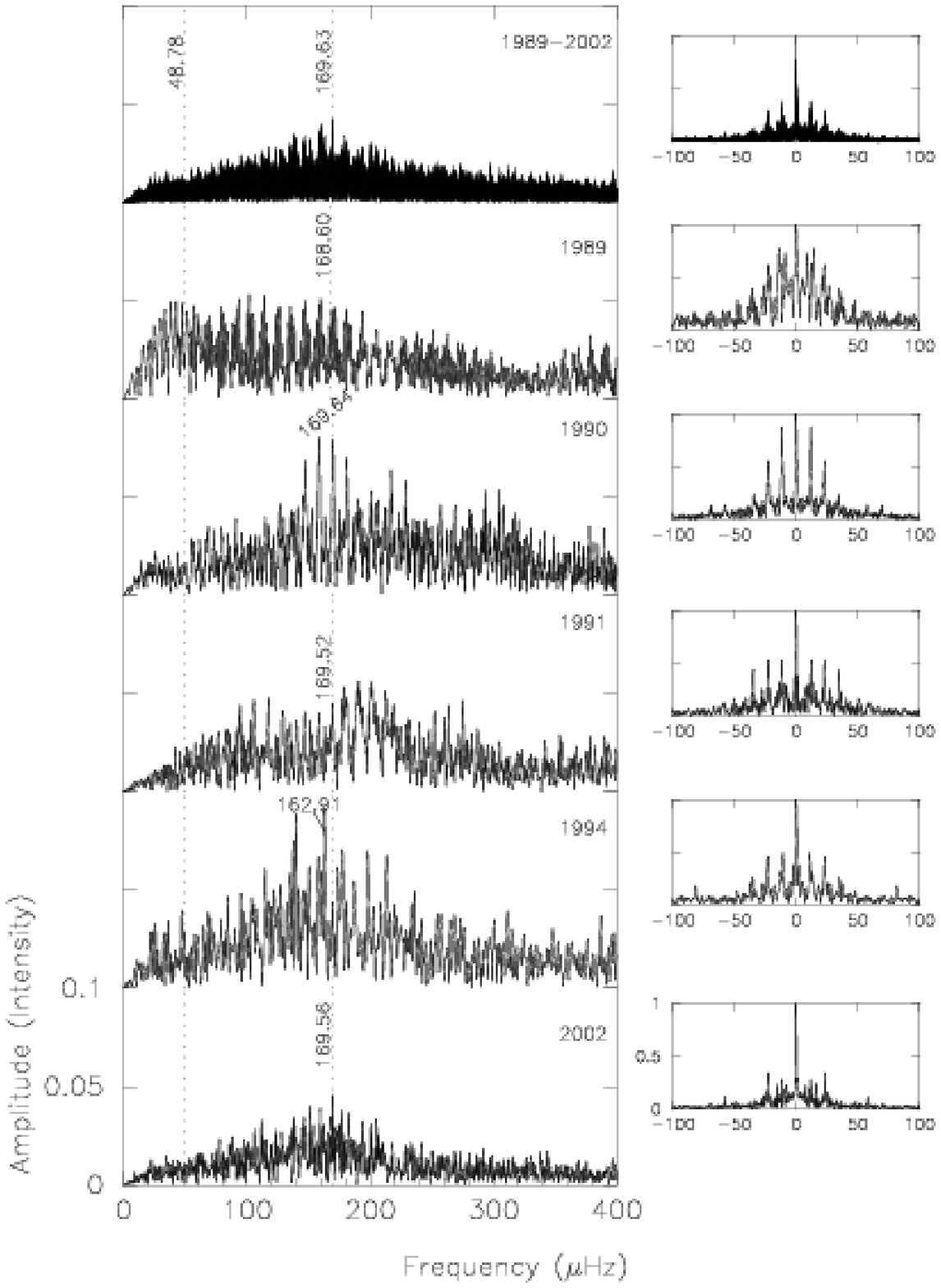}
\caption{\small Discrete Fourier Transform (DFT) amplitude spectra showing light variations in the frequency range: 0 - 400 $\mu$Hz in (from top to bottom) the total combined photometry, the 1989, the 1990, the 1991, 1994 and the 2002 combined photometry.  The frequencies listed are the closest to the QPO peak detected in 2002.  The dotted line on the left marks the location of the orbital period if it was present.  All the plots are on the same scale.  The window spectra are shown on the right also plotted on the same scale.}        
\label{l:archive1}
\end{center}
\end{figure*}
More accurate values of the beat, the harmonic of the beat and the spin frequencies of TX Col were determined from the 1989-2002 combined photometry.  Values of $\omega-\Omega=474.803499\pm0.000089$ $\mu$Hz (2106.13444 $\pm$ 0.00040 s), $\omega=523.584953\pm0.000099$ $\mu$Hz (1909.90974 $\pm$ 0.00036 s) and 2($\omega-\Omega)=949.447975\pm0.000018$ $\mu$Hz (1053.24356 $\pm$ 0.00002 s) were measured.  It should be noted that the errors quoted above are formal estimates from DFTs after fitting by least-squares a sinusoid to the data, and therefore are optimistic.  However, spectral windows show no cycle count ambiguity for the total DFT, suggesting that the periods are stable (this can be seen in Figure~\ref{l:archive1}).

The 1989-2002 combined photometry, however, does not show any modulation at the orbital frequency, and the orbital frequency was determined by taking the difference between the spin and the beat frequencies and was found to be $\Omega = 48.781454\pm0.000013~\mu$ Hz (5.6943317 $\pm$ 0.0000015 h).  The orbital period and the spin period were used to derive the orbital and the spin radial velocity ephemerides, respectively \citep{mhl07c} (hereafter Paper II).

\section{Quasi-Periodic Oscillations}
\label{sec:qpo}
 The 2002 combined photometry (Figure~\ref{l:photplot}, upper panel) shows high-amplitude QPOs with a dominant QPO frequency appearing at $\sim$170 $\mu$Hz.  

To check if this QPO peak was due to noise, the data were subjected to a Fisher Randomisation test \citep{fis35}.  This involves the construction of an artificial dataset of the same mean and the same standard deviation as the original, and the random swapping of the y-data values while the x-data values are kept the same.  The y-values are randomly moved so that they are associated with
different x-points.  Periodograms of the swapped data are then computed (10 000 times in this case) and the height of the resulting noise peaks in the 10 000 periodograms compared with that of the peaks in the original periodogram.  Any peak in the original periodogram with a height less than that in the swapped data is most likely a noise peak and is rejected.  The lower the number of periodograms with higher peaks, the better.  This means that the probability that the peak under examination is a noise peak is $n/10 000$, where $n$ is the number of periodograms with higher peaks.  
Strictly speaking, this is not a confidence level.  This method is nonparametric in a sense that it does not rely on a model specified in terms of a set of unknown parameters.  It just gives an indication of the believability of the peak.  After this exercise it was found that the QPO was likely not due to noise.  

The data were fitted at the QPO frequency on consecutive nights and the results are displayed in Table~\ref{tab:02splqpopar} and in Figure~\ref{l:archivespl}.  As can be seen in Table~\ref{tab:02splqpopar}, the phase of peak maximum of the 170 $\mu$Hz QPO frequency shifts from one night to the next, relative to the first data point of the night of 16 Jan 02 which was chosen as the zero point (since those data possibly have the highest amplitude), confirming that this period is quasi-periodic.
However, the DFTs of the archival data show that the QPO period is also present in the 1990 photometry (Figure~\ref{l:archive1}), and perhaps in the 1994 data, and this suggests that this period is stable on a long timescale and is a QPO that persistently reappears due to some physical/geometrical changes and/or characteristic of TX Col.  The QPO period is also present in the 1989-2002 combined photometry and has the highest amplitude in this dataset (Figure~\ref{l:archive1}).  The QPO frequency was measured from the DFT of the 1989-2002 combined data, and a value of 169.630206 $\pm$ 0.000047 $\mu$Hz (5895.17648 $\pm$ 0.00163 s) was obtained.
\begin{figure}
\begin{center}
\includegraphics[width=90mm]{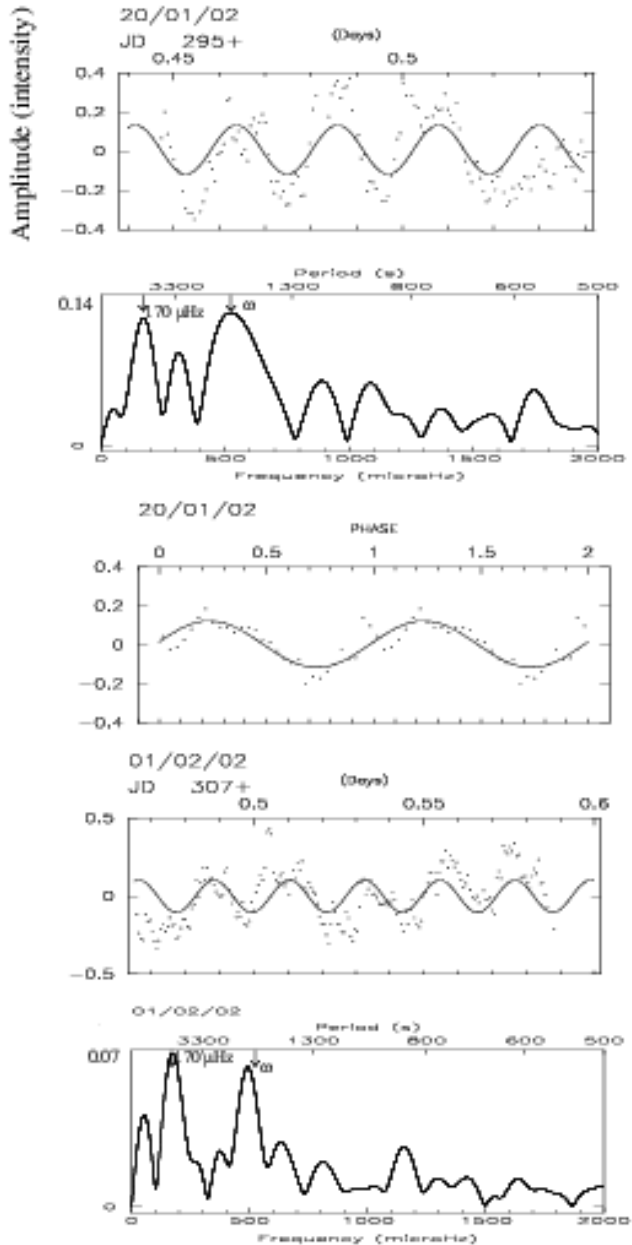}
\caption{\small The top panel shows the light curve of 20 Jan 02 fitted at the spin frequency.  The second panel from the top shows a DFT where the spin and the QPO frequencies were detected, and the middle panel shows the data phase-folded on the spin frequency.  The fourth panel is the light curve of the 01 Feb 02 also fitted at the spin frequency, and the fifth panel is a corresponding DFT.  The solid line represents a fit to the data.}
\label{l:archivespl1}
\end{center}
\end{figure}
\begin{figure}
\begin{center}
\includegraphics[angle=0,width=70mm]{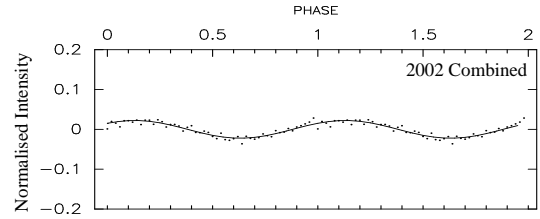}
\caption{\small The simultaneous photometry (2002 combined data set) of TX Col folded on the spin ephemeris, $HJD(maximum) = 2452290.286025 + 0.022105436(4)E$, and binned using 50 bins.  Maximum line intensity is observed at phase $\phi \sim0.14$.  The horizontal scale is in spin phase.}       
\label{rv:48rvel-simphotsp}
\end{center}
\end{figure}

The light curve of 20 Jan 02 (SAAO) (see Figure~\ref{l:archivespl}, middle column of panels, fourth panel from the top) shows an interesting behaviour; excursions or a change in frequency between JD=2452295.44 - 2452295.54 where in one QPO cycle approximately three shorter oscillations, on the timescale of the spin or beat period, are observed.  The light curve of 01 Feb 02 (between JD=2452307.46 - 2452295.6) shows a nearly similar effect.  The data within the above-mentioned JD ranges are strongly modulated near the spin frequency (see first and fourth panels in Figure~\ref{l:archivespl1}) and the DFTs show a peak near the spin frequency (second and last panels in Figure~\ref{l:archivespl1}).  

\section{Spin Variations}  
\label{sec:spinvar}
The data of 20 Jan and 01 Feb falling within the HJD ranges mentioned above, were phase-folded on the radial velocity spin ephemeris $HJD(maximum) = 2452290.286025 + 0.022105436(4)E$ which is derived in Paper II using a spin period determined from the 1989-2002 photometry (Section~\ref{sec:dfts}).  We phased our spectroscopy such that \mrs\ appears at $\phi=0.0$.  The data shows maximum intensity near phase $\phi=0.2$ (middle panel in Figure~\ref{l:archivespl1}).  The data of the 01 Feb 02 (not shown) also showed maximum intensity near phase $\phi=0.2$.
Figure~\ref{rv:48rvel-simphotsp} shows the 2002 combined photometry phase-folded on the radial velocity spin ephemeris (see above).  Maximum intensity is seen at phase $\phi \sim0.14$.

\section{\bf Discussion and Interpretation}

\label{dis:alldisc}

Optical beat modulations are thought to result from reprocessing of X-rays in regions that are fixed in the orbital frame of reference such as the front face of the \snd\ and/or the bright spot \citep{pat81,has81,wsb82}.  The reprocessing model has been used by \cite{buc89a} to explain the optical beat frequency observed in TX Col.  

The disc-overflow model, where beat modulations result from the interaction between the stream of material from the \snd\ rotating with the binary frame at $\Omega$, and the magnetosphere spinning with $\omega$, after the stream has hit and overflowed the outer edge of the disc, has been used successfully as an alternative model to explain X-ray beat pulses.  
It is generally accepted that disc-overflow accretion will result in the simultaneous existence of the beat and the spin pulses in the data, having comparable amplitudes \citep{hel98,nor97}.  These pulses have been observed in the X-rays of TX Col, which establishes disc-overflow as one of the modes of accretion. 
Our optical data of 2002 has shown a dominant modulation at the beat period and another modulation at the spin period.  The spin and the sideband (beat period) is not always detected in TX Col, which is interesting.  This is possibly a result of disc-overflow and will be discussed in detail in Paper II.

In addition to the beat and spin modulations, TX Col amplitude spectra are dominated by high-amplitude QPOs.

We investigate five different models to explain the QPO periodicity.  

(1) A successful model for QPOs and dwarf nova oscillations (DNOs) was proposed by \cite{war02} where QPOs are caused by slow-moving prograde waves at the inner edge of the disc.  \cite{war03} showed that many observations in CVs and X-ray binaries obey the relation $P_{QPO}/P_{spin}\sim15$.  This model explained the QPOs observed in the IP GK Per, where $P_{QPO}/P_{spin}\sim14$ \citep{hl94}.  However, TX Col does not obey this relation since $P_{QPO}/P_{spin}\sim3$ and so this model cannot be applied as it is to this system.

(2) \cite{ret04} reported evidence for large superhumps in TX Col, at 7.1 h (positive superhump) and at 5.2 h (negative superhump), in addition to the orbital period.  These periods are understood as resulting from beating of the orbital period and the apsidal or modal precession of the disc.  Their observations of TX Col taken between December 2002 and February 2003, about a year after our campaign, showed large-amplitude QPOs.  
A possibility, therefore, is that the interaction between the superhump frequency and the Keplerian frequency of the material at the outer disc produces a new frequency - the QPO.  Using the orbital parameters of TX Col we find that the Keplerian period of the material at the outer disc edge is

\begin{equation}
P_{KEP}=\frac{2\pi R_{out}}{v_{KEP}} \sim 2000-12000 \hspace{0.1cm} {\rm s},
\label{eq:one} 
\end{equation}
for any reasonable values of $v_{KEP}$ between $\sim$400-600 \ks\ ($V_{KEP}{\rm sin}i\sim172$ \ks\ - $i<25$\zdg) and of the outer disc radius \[R_{out}=\frac{GM_{1}}{v_{KEP}^{2}} \sim 2-8\times10^{10} \hspace{0.1cm} {\rm cm.} \] The \wtd\ mass range between M$_{1} \sim0.5 - 1$ M$_{\odot}$ \citep{ram00,sul05} is considered here.

\[ \frac{1}{P_{KEP}}- \frac{1}{P_{SH}}=\frac{1}{P_{QPO}} \] gives QPO periods that we observe in the data ($\sim$6000 s) for values of $P_{KEP}$ near 5000 s.  This would imply a smaller disc, though.

The presence of the 7.1 h period is deemed unlikely, whereas that of the 5.2 h is possible but not conclusive (Tansel AK;private communication).  Our extensive data do not show evidence for these superhump periods.

(3) The theoretical analysis of \cite{kin93} and \cite{wyn95} suggested that the flow of matter in IPs can take the form of diamagnetic 'blobs' that orbit about the \wtd.  \cite{hel02} argued that, following the theory of \cite{kin93} and \cite{wyn95}, TX Col can be explained as having a combination of a stream and orbiting blobs.
A similar suggestion was put forward for GK Per where it was thought that QPOs result due to vertically extended 'blobs' orbiting within the inner accretion disc edge and providing modulated reprocessing of, or illumination by, the \wtd\ \citep{mor96}.

We find that the Keplerian period of the material at the inner disc edge is $\sim 200-600$ s, for any \wtd\ mass between M$_{1} \sim0.5 - 1$ M$_{\odot}$ and inner disc radius, \[R_{in}= \frac{GM_{1}}{v_{KEP}^{2}} \sim 2-4\times10^{9} \hspace{0.1cm} {\rm cm.}\] These periods are inconsistent with the QPO time-scales of $\sim6000$ s observed in our data.  Therefore theories where the QPO is a beat between the spin frequency and the frequency of material orbiting the \wtd\ at the inner edge of the disc or where the QPO results from reprocessing off blobs or bulge orbiting at the inner edge of the disc \citep{wat85} are not supported by our observations for TX Col.  

(4) However, the beat of the spin period with the Keplerian period at the outer disc i.e. 
\[ \frac{1}{P_{spin}}- \frac{1}{P_{KEP}}=\frac{1}{P_{QPO}} \] 
gives $P_{QPO}\sim6000$ s which we observe in our data, for values of $P_{KEP}$ in the lower range near 3000 s (Equation~\ref{eq:one}) and for reasonable values of $R_{out}\sim3\times10^{10}$ \ks\ \citep{buc89} and $v_{KEP}\sim600$ \ks).
Though this model seems to give the expected result, it alone does not explain why the QPO variation has a higher amplitude (compared to the beat and the spin periods).

(5) Therefore we suggest that in addition to there being `blobs' at the outer edge of the disc from which \wtd\ emission is reprocessed to give rise to QPO frequency, there is modulated accretion occuring at the magnetosphere/disc boundary that gives rise to the same QPO frequency.

\cite{spr93} showed that conditions at the inner edge of the disc can cause variations of the magnetosphere boundary and that material can accumulate outside the magnetosphere.  
\cite{spr93} pointed out that their model could be applied to IPs to explain the QPO phenomena seen in these systems.  This model was used recently by \cite{mhl07b} to describe the outburst of EX Hya.  

Our results have shown that maximum intensity of the continuum light occurs at spin phase $\sim0.2$, when the narrowing `neck' of the accretion curtain is nearly facing the observer (Paper II).  Since the continuum light curves are dominated by the QPOs, it follows that most of the QPO emission comes from this region, near the \wtd.  The spin modulation appearing in the QPO continuum light curves also shows maximum intensity near this phase ($\sim0.2$; Section~\ref{sec:qpo}), suggesting that continuum spin modulations also emanate from this region. 
The variable intensity and excursions in the QPO light curves (Figure~\ref{l:archivespl}) suggest that it is an accretion process that gives rise to the QPO emission.  We proposed that it is near the above-mentioned region where the QPO modulations result, due to accretion.  

Between JD=2452295.44 - 2452295.54 and JD=2452307.46 - 2452295.6) there are possibly no `blobs' that are picked up by the accretion curtains and accreted via the Spruit and Taam mechanism by the \wtd.  This results in the observed spin modulated emission in the QPO continuum light curves.  

We suggest that the material that forms a `base excursion' (Paper II; see also \cite{hel89,mhl07b}) due to overflow stream falling near the magnetosphere/disc boundary, and the `blobs' that drift from the outer disc towards this same shocked region, pile up near this region and are dumped onto the surface of the \wtd\ via a mechanism similar to that of Spruit and Taam before the field lines snap to produce a prograde travelling wave (or `wall') of \cite{war02}.  

The critical density required to push the magnetosphere inward for the accretion of the accumulated `blobs' to take place is possibly reached quicker in TX Col than in EX Hya, resulting in the frequent accretion of the `blobs' and in the production of the QPOs that we observe in the data.  This could explain why we do not see outbursts in TX Col.  

The viscous time scale at the corotation radius, r$_{co}$, predicted by the \cite{spr93} model can roughly be estimated to be $t_{0}=1/\bar{\nu_{0}}\Omega_{s}\sim356$ s \citep{spr93}, where $\bar{\nu_{0}}=\alpha(\frac{H}{r_{co}})^{2} \sim0.1$ and assuming the $\alpha$ viscosity parameter is $\sim0.1$ \citep{sha73}.  These time scales are inconsistent with the observed QPO time scales.  However, at R$_{out}$ where we suggest there are orbiting `blobs', $t_{0}\sim5000$ s.
The latter time scales are consistent with the QPO time scales.  This could suggest that there is evolution of `blobs' from R$_{out}$ toward the magnetosphere.  

This could also suggest that TX Col has extended accretion curtains where material is accreted from a ring near the Roche lobe, a similar situation as in EX Hya \citep{kin99,bel02,nor04,mhl07a}.  In this geometry the QPO period would result due to the `blobs' orbiting in the ring of material being swept up by the magnetic field lines.  This would occur when an orbiting `blob' is on the side facing the magnetic field lines.
This is unlikely, though, given the $P_{spin}/P_{orb}$ ratio of TX Col.  Also, such a behaviour can be confirmed by the detection of a spin period modulated at radial velocities near those of the outer ring material due to corotation of outer ring material with the accretion curtain \citep{mhl07a}.
\section{Summary}
The photometry of TX Col has been dominated by QPOs but no interpretation for their origin had been provided before.  A 5900 s QPO period is detected in the 1990, 1994 and 2002 photometry and we interprete it as follows: the QPO period results due to the beating of the Keplerian period of the orbiting `blobs' with the spin period and from the storage and release of `blobs' near the magnetosphere, where the stored material is rapidly accreted by the \wtd.  
\section*{Acknowledgments}

{NM would like to acknowledge financial support from the Sainsbury/Linsbury Fellowship Trust and the University of Cape Town.  NM would  also like to thank D. O'Donoghue and Tom Marsh for the use of their programs, Eagle and Molly, respectively.}

\addcontentsline{toc}{chapter}{References}
\renewcommand{\bibname}{References}

\end{document}